\documentclass[a4paper,11pt]{article}
\usepackage{pos}
\usepackage{hyperref}
\usepackage{orcidlink}
\def\kt{\ensuremath{k_{\rm T}}}
\def\zm{\ensuremath{z_{\rm M}}}
\def\cascade{{\sc Cascade}}

\input epsf.tex
\def\desepsf(#1 width #2){\epsfxsize=#2 \epsfbox{#1}}
\def\kt{\ensuremath{k_{\rm T}}}

\def\qt{\ensuremath{q_{\rm t}}}
\def\zdyn{\ensuremath{z_{\rm dyn}}}
\def\ptll{\ensuremath{p_{\rm T}(\ell\ell)}}

\newcommand{\mdy}{\ensuremath{m_{\text{DY}}}}

\newcommand{\PBM}{PB}

\newcommand{\sqrts}{\ensuremath{\sqrt{s}}}

\title{Recent progress in transverse momentum dependent (TMD) Parton Densities and corresponding parton showers}

\author*{S.~Taheri~Monfared\orcidlink{0000-0003-2988-7859}}

\affiliation{Deutsches Elektronen-Synchrotron DESY, Germany}

\emailAdd{taheri@mail.desy.de}

\abstract{
The parton branching method is crucial for Monte Carlo generators, which are essential for high-energy physics predictions. We examine the impact of soft gluons on inclusive collinear and Transverse Momentum Dependent (TMD) parton densities. By applying the Parton-Branching (\PBM) method, we identify the non-perturbative Sudakov form factor with the integration range $z \to 1$, which is often neglected in collinear parton shower approaches.

The significance of soft gluons is demonstrated through the transverse momentum spectrum of Drell-Yan lepton pairs, resulting in an intrinsic-\kt\ distribution width that remains independent of \sqrts, contrary to observations in collinear parton shower approaches. This behavior is attributed to the non-perturbative Sudakov form factor.

}

\FullConference{31st International Workshop on Deep Inelastic Scattering (DIS2024)\\
 8–12 April 2024\\
Grenoble, France\\}

\begin{document}
\maketitle

\section{Introduction}
Experimental measurements in high-energy particle physics are often more precise than the theoretical predictions generated by Monte Carlo (MC) event generators. 
One source of uncertainty in collinear parton shower is neglecting soft gluon emissions in backward evolution. Another potential problem comes from the differences between the details in forward evolution used for Parton Distribution Functions (PDFs) extraction and backward evolution in the parton shower algorithms. This seperation of forward and backward evolution, gives the freedom in choosing parameters of parton showers.

The Parton Branching (\PBM) method serves as a straightforward connection between the Collins-Soper-Sterman (CSS) approach and parton shower methods in MC event generators. It allows for a detailed examination of their connection, especially in understanding the origins of the non-perturbative Sudakov form factor \cite{Bubanja:2024puv,Mendizabal:2023mel}.

\section{Parton Branching method}

The \PBM\ method is a versatile MC approach for generating QCD high-energy predictions using Transverse Momentum Dependent (TMD) PDFs, commonly referred to as TMDs. A key component of this method is the forward evolution equation \cite{Hautmann:2017xtx,Hautmann:2017fcj}, which describes the evolution of parton density through real, resolvable branchings and virtual, non-resolvable contributions, managed using Sudakov form factors. The \PBM\ method's evolution equation for a TMD density ${\cal A}_a(x,{\bf k}, \mu^2)$ for parton $a$ at scale $\mu$ can be expressed in integral form as:
\begin{eqnarray}
\label{evoleqforA}
{\cal A}_a(x,{\bf k}, \mu^2) 
&=&  
\Delta_a(\mu^2) \, {\cal A}_a(x,{\bf k}, \mu^2_0)  
+ \sum_b \int_{\mu^2_0}^{\mu^2} \frac{d^2 {\bf q}^{\prime}}{\pi {\bf q}^{\prime 2}} \, \frac{\Delta_a(\mu^2)}{\Delta_a({\bf q}^{\prime 2})} \nonumber\\
&& \times \int_x^{\zm} \frac{dz}{z} \, P_{ab}^{(R)}(\alpha_s, z) \, {\cal A}_b\left(\frac{x}{z}, {\bf k} + (1-z) {\bf q}^{\prime}, {\bf q}^{\prime 2}\right),     
\end{eqnarray}
where $x$ denotes the longitudinal momentum fraction, ${\bf k}$ represents the 2-dimensional transverse momentum vector with $\kt = |{\bf k}|$, and $|{\bf q}^{\prime}| = q^\prime$. The initial distribution ${\cal A}_a(x,{\bf k}, \mu^2_0)$ in equation (\ref{evoleqforA}) at scale $\mu_0$ is parameterized in terms of a collinear parton density at the starting scale and an intrinsic-$\kt$ distribution modeled as a Gaussian function with width $\sigma$. The width parameter $\sigma$ of the Gaussian distribution is related to $q_s$ by $q_s = \sqrt{2} \sigma$.

 The Sudakov form factor, $\Delta_a(\mu^2)$, derived from the momentum sum rule and unitarity, represents real, resolvable splittings as a non-emission probability. This offers a clear, intuitive view of evolution as a sequence of branchings, enabling the PB evolution equation to be solved using MC techniques through a parton branching algorithm.
Beyond the evolution equation, the \PBM\ method includes a procedure for fitting initial distribution parameters to experimental data via the xFitter platform \cite{Alekhin:2014irh}. The resulting PB TMDs and PDFs \cite{BermudezMartinez:2018fsv,Jung:2021vym,Jung:2021mox} are accessible through TMDlib \cite{Abdulov:2021ivr} and in LHAPDF  format for use in TMD MC generators. An important generator is the TMD MC generator \cascade\ \cite{CASCADE:2021bxe}, which implements the TMD initial state parton shower with backward evolution guided by PB TMDs. 
In \cascade\ we have no inconsistencies from the differences between the details in forward evolution used for PDFs extraction and backward evolution in the parton shower algorithms. 

\subsection{Introducing non-perturbative Sudakov form factor}
The Sudakov form factor $\Delta_a ( \mu^2 )$ is essential in the formulation of the \PBM\ method:
\begin{equation}
\label{sud-def}
 \Delta_a ( \mu^2  ) = 
\exp \left(  -  \sum_b  
\int^{\mu^2}_{\mu^2_0} 
{{d {\bf q}^{\prime 2} } 
\over {\bf q}^{\prime 2} } 
 \int_0^{\zm} dz \  z 
\ P_{ba}^{(R)}\left(\alpha_s , 
 z \right) 
\right) 
  \;\; ,   
\end{equation}
The resolvable splitting functions $P_{ba}^{(R)}(\alpha_s, z)$ describe the process where a parton $a$ splits into a parton $b$, with $z$ representing the ratio of their longitudinal momenta. These functions typically correspond to DGLAP splitting functions, either at leading or next-to-leading order. To ensure numerical stability during calculations, a parameter $z_M$ is introduced, where $z_M = 1 - \epsilon$ and $\epsilon \to 0$. It is crucial that $\epsilon \to 0$ to facilitate accurate cancellation of terms in the evolution equation derivation and to accurately reproduce the DGLAP limit, as detailed in Ref.~\cite{Hautmann:2017fcj,Hautmann:2017xtx}. This approach also ensures stable solutions for TMD distributions.

Having angular ordering condition in the evolution, one can define a seperation tool of $\zdyn = 1 - q_0/q'$. The Sudakov form factor can now be decomposed into perturbative ($0 < z < \zdyn$) and non-perturbative ($\zdyn < z < \zm$) parts \cite{Mendizabal:2023mel,Martinez:2024twn}:
\begin{equation}
\label{eq:divided_sud}
\Delta_a(\mu^2) = \Delta_a^{(\text{P})}(\mu^2) \cdot \Delta_a^{(\text{NP})}(\mu^2),
\end{equation}
where, in $\Delta_a^{(\text{P})}$, the integral over $z$ ranges from $0$ to $\zdyn$, while in $\Delta_a^{(\text{NP})}$, it ranges from $\zdyn$ to $\zm$.

The \PBM\ method facilitates a connection to the CSS formalism: by implementing angular ordering with $\zdyn = 1 - q_0/q'$ and utilizing the transverse momentum $\qt = q' (1-z)$ as a scale in $\alpha_s$, it becomes feasible to derive purturbative components of the CSS Sudakov factor, extending up to next-to-next-to-leading order, as elaborated in Ref.~\cite{Martinez:2024twn}.
Moreover, a significant advantage of the \PBM\ method is the ability to compute the non-perturbative Sudakov form factor $\Delta_a^{(\text{NP})}$, which can be determined through fitting inclusive distributions~\cite{BermudezMartinez:2018fsv}. In contrast, the CSS approach requires additional constraints on this form factor from exclusive measurements.

\subsection{Non-perturbative Sudakov form factor in inclusive and exclusive distributions}

In Ref. \cite{Mendizabal:2023mel}, it was demonstrated that non-perturbative Sudakov form factors play a crucial role in inclusive distributions, such as collinear parton densities and Drell-Yan transverse momentum spectra. These soft emissions are integral to the $\overline{MS}$-scheme, as neglecting them would result in the non-cancellation of important singular terms.

\begin{figure}[htb]
\centerline{
\includegraphics[width=0.6\textwidth]{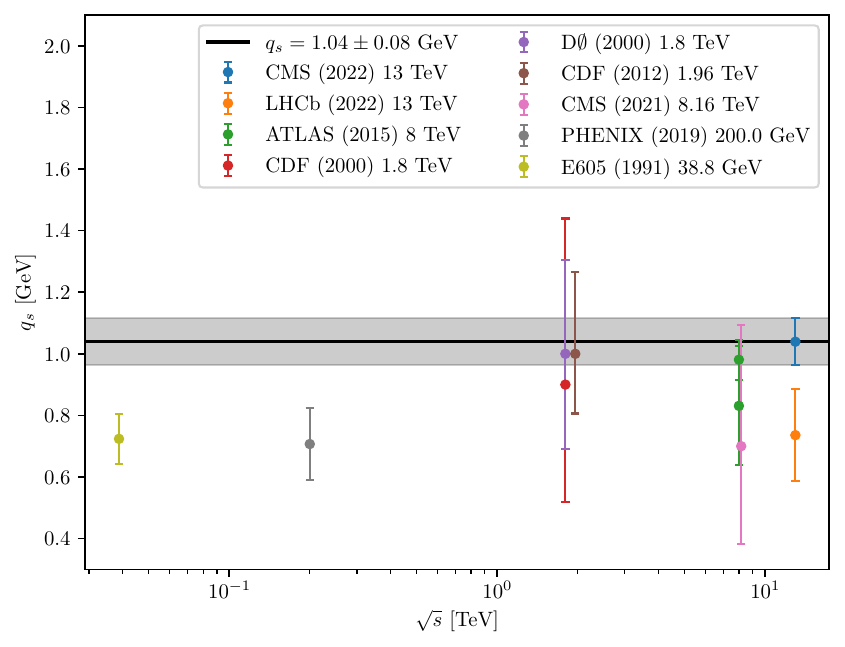}
}
\caption{The width parameter $q_s$ of the intrinsic-\kt\ distribution as a function of the center-of-mass energy \sqrts . (Plot taken from Ref.~\cite{Bubanja:2023nrd}) }
\label{PB-kt}
\end{figure}

Ref. \cite{Bubanja:2023nrd} focuses on the transverse momentum spectrum $\ptll$ and determines the width of the intrinsic-$\kt$ distribution through precise measurements at LHC energies \cite{CMS:2022ubq} across a wide range of DY masses $\mdy$. Lower-energy measurements were also examined, showing that all data can be effectively described using the \PBM\ approach. The width of the intrinsic-$\kt$ distribution exhibits a very mild dependence on the center-of-mass energy $\sqrt{s}$, as depicted in Fig.~\ref{PB-kt}.

\begin{figure}[htb]
\centerline{
\includegraphics[width=0.6\textwidth]{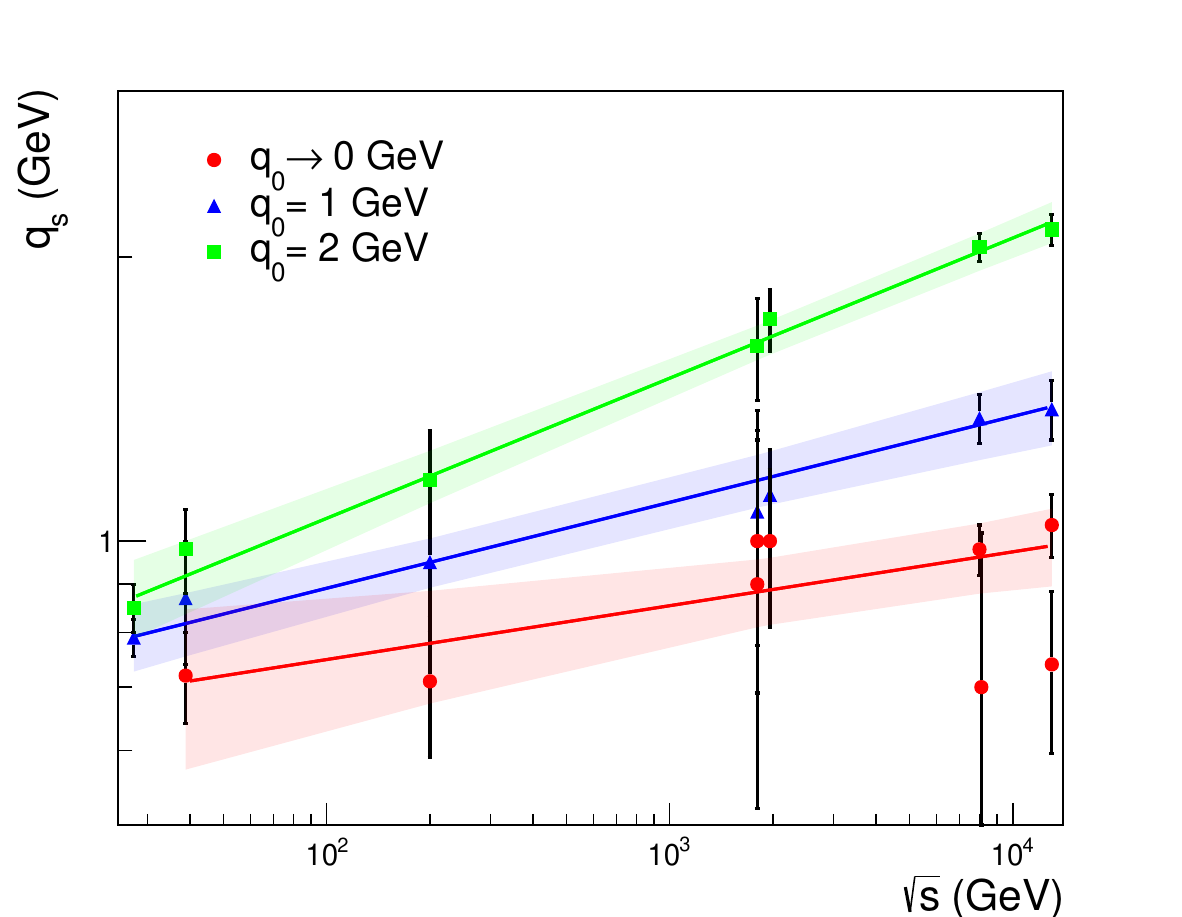}
}
\caption{The width parameter $q_s$ of the intrinsic-\kt\ distribution as a function of center-of-mass energy \sqrts . The uncertainty bands show the 95\% CL. 
 (Plot taken from Ref.~\cite{Bubanja:2024puv}) }
 
 \label{kt-width}
\end{figure}

In Ref. \cite{Bubanja:2024puv}, it was confirmed that this stable result is obtained with the original \PBM\ set, where $q_0 < 0.01$ GeV. As shown in Fig.~\ref{kt-width}, excluding the non-perturbative Sudakov form factor reveals a strong dependence of the width parameter $q_s$ on the center-of-mass energy $\sqrt{s}$.

It was also established in Ref. \cite{Mendizabal:2023mel} that these soft emissions have essentially no effect on final state hadron spectra and jets.

\section{Conclusion}
The \PBM\ approach bridges analytic resummation via the DGLAP equation for inclusive parton distributions with TMD resummation and parton shower methods. It accurately reproduces semi-analytical solutions of the DGLAP equation when integrating over soft gluons ($z \to 1$), crucial for canceling terms in NLO cross sections. \PBM\ directly provides TMD distributions by simulating each branching and including kinematic relations.

Non-perturbative, soft gluons are crucial for describing the low-$\ptll$ Drell-Yan spectrum. \PBM\ reveals an $\sqrt{s}$-independent width of the intrinsic-\kt\ distribution, contrasting collinear shower simulations. Soft gluons affect inclusive distributions but not observable hadron spectra.

\section*{Acknowledgments}
The results discussed in this contribution are based on the work in Refs. \cite{Bubanja:2024puv,Mendizabal:2023mel,Bubanja:2023nrd}. Many thanks to all co-authors for collaboration. I am grateful to the organizers of DIS 2024 for the invitation to present these results at the workshop.

\end{document}